\documentclass[onecolumn]{simplejournal}

\usepackage{orcidlink}

\title{Ontology-based knowledge graph infrastructure for interoperable atomistic simulation data}

\author{
  Abril Az\'{o}car Guzm\'{a}n$^{1}$\email{a.azocar.guzman@fz-juelich.de}\orcidlink{0000-0001-7564-7990},
  Sarath Menon$^{2}$\orcidlink{0000-0002-6776-1213},
  Tilmann Hickel$^{3}$\orcidlink{0000-0003-0698-4891},
  and Stefan Sandfeld$^{1}$\orcidlink{0000-0001-9560-4728}\\[6pt]
  \affil{$^1$Institute for Advanced Simulations -- Materials Data Science and Informatics (IAS-9), Forschungszentrum J\"{u}lich GmbH, J\"{u}lich, Germany}\\
  \affil{$^2$Interdisciplinary Centre for Advanced Materials Simulation (ICAMS), Ruhr University Bochum, Bochum, Germany}
  \affil{$^3$Bundesanstalt f\"{u}r Materialforschung und -pr\"{u}fung (BAM), Berlin, Germany}
}

\date{\today}

\begin{document}

\maketitle

\begin{abstract}
The reuse of atomistic simulation data is often limited by heterogeneous formats, incomplete metadata, and a lack of standardized representations of workflows and provenance. Here we present an ontology-based infrastructure for representing and integrating atomistic simulation data as a knowledge graph. The approach combines domain ontologies with a software framework that enables data capture both from existing datasets and directly from simulation workflows at the point of generation. Heterogeneous data from multiple sources are normalized into a common, ontology-aligned representation, enabling consistent querying and analysis across datasets. We demonstrate these capabilities through the integration of grain boundary data, cross-dataset analysis of material properties, and extraction of derived thermodynamic quantities from existing simulations. In addition, workflows are represented in a machine-readable form, enabling both forward provenance tracking and partial reconstruction of computational procedures. The resulting knowledge graph contains over 750,000 triples describing nearly 8,000 computational samples. This work provides a practical framework for improving the findability, interoperability, and reuse of atomistic simulation data.
\end{abstract}


\section{Introduction}

Data-driven approaches in materials science increasingly rely on the aggregation and reuse of heterogeneous data from simulations and experiments across a wide range of compositions, structures, and thermodynamic conditions. In this context, atomistic simulations constitute a major source of computational materials data. Methods such as density functional theory and molecular dynamics are routinely used to investigate the structure, energetics, and properties of materials at the atomic scale, and the resulting data are increasingly reused for benchmarking, validation, and the development of machine-learning and artificial-intelligence approaches \cite{BenMahmoud2024, Himanen2019, dePablo2019}.

A major obstacle to reuse is that atomistic simulation data are commonly stored in software-specific formats, which limits interoperability across codes and platforms. In addition, metadata are often recorded inconsistently, workflow and provenance descriptions lack standardization, and important simulation parameters may be incompletely or only implicitly documented. As a result, the interpretation and comparison of calculated materials properties often require substantial manual effort. Recent perspectives on accelerated materials discovery and metadata practices for simulation workflows have further emphasized the need for structured, machine-readable representations that support interoperability, reproducibility, and reuse across heterogeneous computational environments \cite{Pyzer-Knapp2022, Horsch2020, Villamar2025, Bonomi2019}.

To address parts of these challenges, a range of software environments and workflow management systems have been developed for computational materials science, including AiiDA \cite{Huber2020}, pyiron \cite{Janssen2019}, and jobflow \cite{Rosen2024}. These tools enable the automation, orchestration, and documentation of computational workflows involving multiple calculations. As studies increasingly span large configurational spaces and complex processing pipelines, the resulting workflows become correspondingly intricate, particularly when considering systems with atomic-scale defects \cite{Walsh2024}. While such frameworks improve reproducibility and workflow management within their respective ecosystems, they do not by themselves provide a shared, machine-actionable representation of workflows, methods, and provenance across different platforms. Recent efforts to define exchangeable workflow representations further highlight that interoperability between workflow systems remains an open challenge, even within modern computational materials infrastructures \cite{Janssen2025}.

In parallel, the materials modelling community has made considerable progress in addressing these challenges through the development of curated databases of structure--property data, such as the Materials Project \cite{Jain2013}, AFLOW \cite{Curtarolo2012}, and NOMAD \cite{Draxl2019}. These infrastructures have substantially improved access to computational materials data and enabled large-scale data-driven studies. However, they have been developed primarily for bulk materials, where structures and properties are more readily standardized. In contrast, defect-containing systems are inherently more complex, as their description depends on local atomic environments, chemical composition, and the specific simulation workflow used to generate them. As a result, defect structures and their associated simulation workflows are much less systematically represented and more difficult to compare across datasets. Moreover, while these databases improve data accessibility, they do not necessarily provide a consistent semantic representation of workflows, methods, and material properties, which limits interoperability and reuse.

To enable meaningful reuse and interpretation of atomistic simulation results, metadata must be recorded in a consistent and structured manner at every stage of the workflow, from the initial atomic structure to derived material properties at larger scales. This requires not only the availability of data, but also sufficient contextual information to describe how results were generated, processed, and related. These requirements are closely aligned with the FAIR principles, which emphasize that research data should be findable, accessible, interoperable, and reusable \cite{Wilkinson2016}. In materials science, the adoption of FAIR principles has therefore become a central objective, requiring not only open access to data but also shared metadata standards that support interoperability and reuse \cite{Scheffler2022, Ghiringhelli2023}. While substantial progress has been made in establishing materials data infrastructures and standards \cite{Butler2024}, simulation data remain highly heterogeneous and difficult to integrate across workflows, methods, and software environments. In particular, workflows, methods, provenance, and detailed materials descriptions are often not represented in a machine-interpretable form, which limits integration and reuse. This motivates the need for more expressive digital knowledge representations that capture the full scientific context of atomistic simulation data \cite{Bayerlein2022}.

Motivated by these requirements, semantic approaches have gained increasing attention in materials science as a means to improve interoperability and data reuse \cite{Valdestilhas2023}. Existing efforts include domain ontologies for formalizing materials knowledge and enabling semantic interoperability across heterogeneous data sources, such as the Materials Design Ontology \cite{Li2020} and the PMD Core Ontology \cite{Bayerlein2024}, as well as upper-level frameworks like the Elementary Multiperspective Material Ontology \cite{DelNostro2024}. In parallel, standards such as OPTIMADE have significantly improved programmatic access to and exchange of materials data \cite{Evans2024}, while knowledge-graph-based approaches, including propnet, demonstrate how structured graph representations can capture relationships between materials entities and properties \cite{Mrdjenovich2020}. These developments highlight the potential of semantic artefacts to support machine-actionable materials data; however, they primarily address data exchange, broad domain integration, or property relationships, and do not yet provide a dedicated semantic representation of atomistic simulation workflows, methods, properties, and defect structures within a unified framework.

Our aim is to extend data-driven materials science to more complex systems, particularly materials containing crystallographic defects. To achieve this, we introduce a modular ontology-based framework for the semantic representation of atomistic simulation data, covering material and defect structures, simulation workflows, calculated properties, and their interrelations. We further provide a software stack for constructing materials knowledge graphs from such data. In this framework, the ontologies supply the shared semantic schema, while the knowledge graph realizes this schema as a linked, queryable representation of simulation data and provenance. Together, these components provide an interoperable, machine-actionable, and FAIR foundation for integrating and reusing atomistic simulation data across heterogeneous computational workflows and data sources.

\section{Methods}

\subsection{Ontologies for materials simulation} \label{sec:onto}

In this section, we introduce the ontology framework that forms the semantic foundation of our infrastructure. We develop the Computational Materials Sample Ontology (CMSO) and the Atomistic Simulation Methods Ontology (ASMO), which together provide a machine-actionable representation of material structures and atomistic simulation workflows.

\subsubsection{Computational Materials Sample Ontology}

CMSO describes computational materials science samples, that is, material structures represented in simulations, and provides a formal vocabulary for representing material samples from the atomic scale to the macroscale together with their crystallographic, structural, and compositional information, including crystalline defects.

It is designed for use in the context of atomistic simulations and materials databases. The central concept is the \texttt{ComputationalSample}, with \texttt{AtomicScaleSample} as a key subclass. The main concepts used to describe structures can be grouped into the following categories: materials, microstructure, crystal structure, atoms and chemistry, simulation cell, geometry, and links to defect ontologies.

The ontology follows a modular approach in which extensions are introduced according to the length scale of the computational method. These include \texttt{NanoscaleSample}, \texttt{MesoscaleSample}, \texttt{MicroscaleSample}, and \texttt{MacroscaleSample}, each of which can be extended with scale-specific descriptions.

\begin{figure}[ht!]
\centering
\includegraphics[scale=0.08]{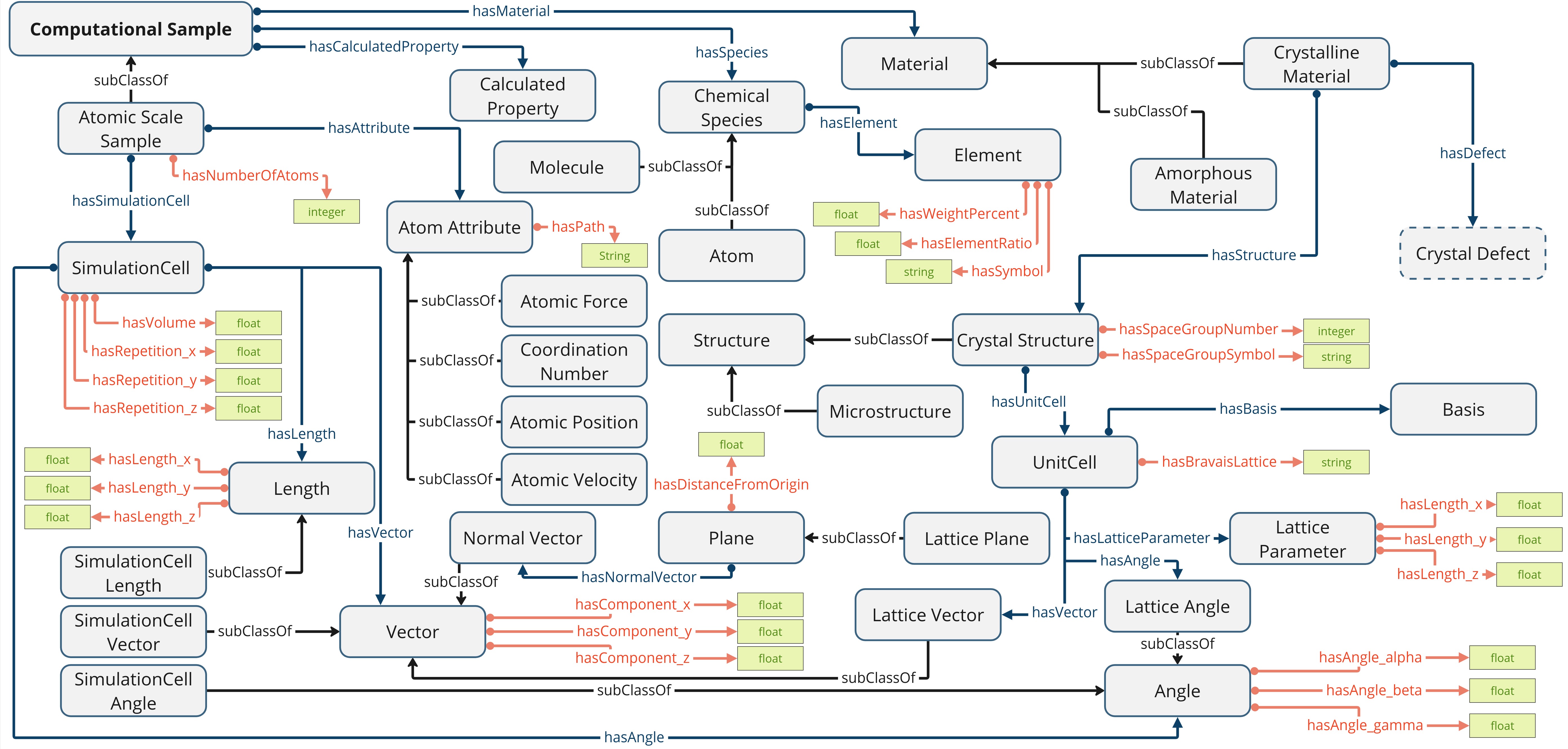}
\caption{CMSO Ontology.} 
\label{fig:cmso}
\end{figure}

The ontology contains 46 classes, 20 object properties, and 33 data properties. An overview of CMSO is shown in Fig.~\ref{fig:cmso}. The ontology is available through its GitHub repository \cite{cmso_git} and archived on Zenodo \cite{cmso_zenodo}.

\subsubsection{Atomistic Simulation Methods Ontology}

ASMO provides a formal vocabulary for describing computational methods used in atomistic materials simulations, together with the simulation workflows through which data are generated.

The ontology is structured around the core class \texttt{Simulation}, which is further characterized by \texttt{ComputationalMethod}, \texttt{SimulationAlgorithm}, and \texttt{SimulationParameter}. It is specialized into major method families such as \texttt{DensityFunctionalTheory}, \texttt{MolecularDynamics},\\ \texttt{MolecularStatics}, \texttt{KineticMonteCarloMethod}, and \texttt{AbInitioMolecularDynamics}. Its result space is centered on \texttt{CalculatedProperty} and \texttt{PhysicalQuantity}, which organize general concepts such as \texttt{Energy}, \texttt{Force}, \texttt{Length}, \texttt{Mass}, \texttt{Pressure}, \texttt{Stress}, \texttt{Temperature}, \texttt{Time}, and \texttt{Volume}, together with more specific outputs including \texttt{BulkModulus}, \texttt{ShearModulus}, \texttt{YoungsModulus}, \texttt{PoissonsRatio}, \texttt{FormationEnergy}, \texttt{TotalEnergy}, \texttt{VirialPressure}, and \texttt{TotalMagneticMoment}. ASMO also includes dedicated branches for \texttt{StructureManipulation}, \texttt{SpatialTransformation}, \texttt{PointDefectCreation}, \texttt{InteratomicPotential}, \texttt{StatisticalEnsemble}, and\\ \texttt{MathematicalOperation}.

\begin{figure}[ht!]
\centering
\includegraphics[scale=0.08]{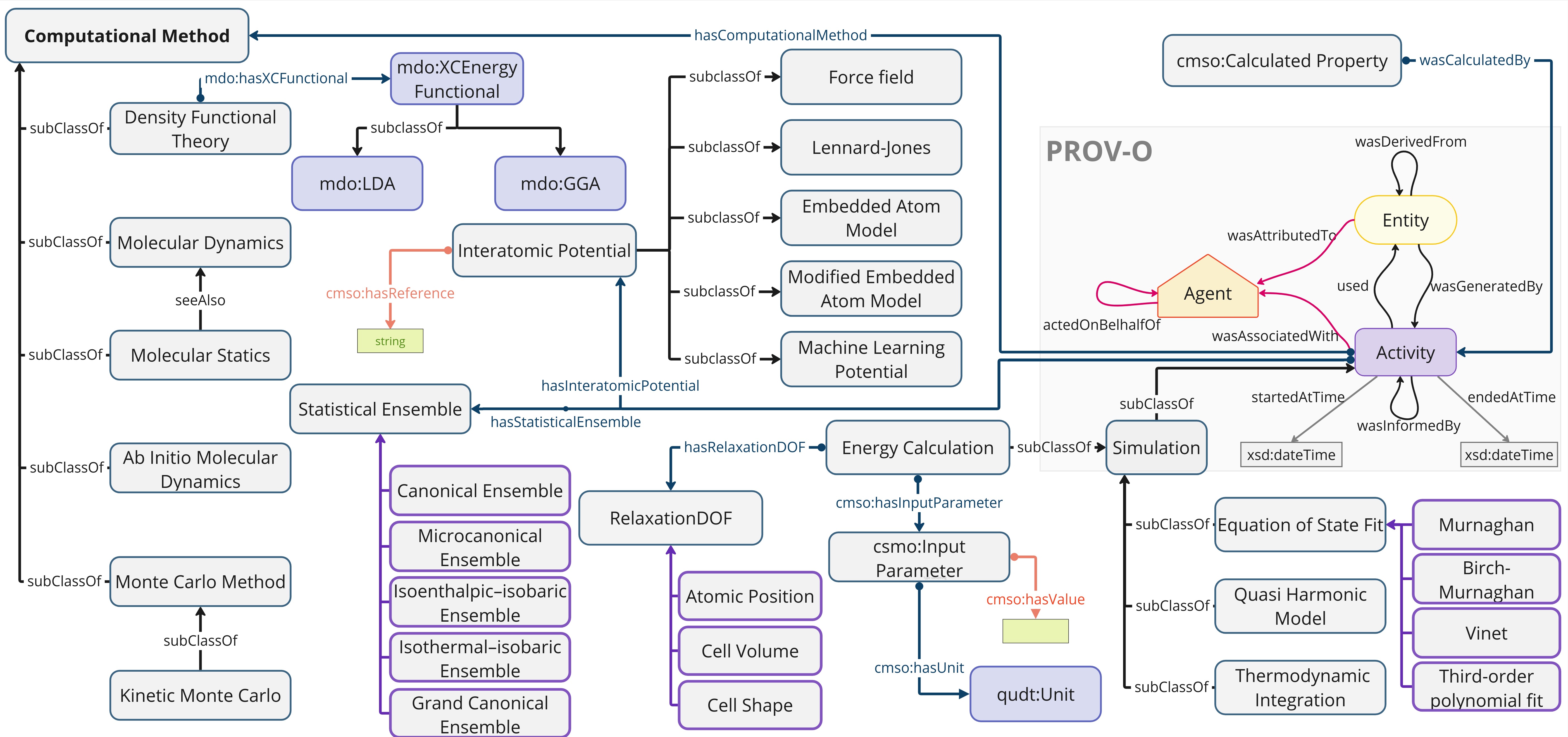}
\caption{ASMO Ontology.} 
\label{fig:asmo}
\end{figure}

To represent simulation workflows and their provenance, ASMO builds on the W3C PROV-O provenance model. The ontology contains 105 classes, 25 object properties, 16 data properties, and 30 named individuals. An overview of ASMO is shown in Fig.~\ref{fig:asmo}. The ontology is available through its GitHub repository \cite{asmo_git} and archived on Zenodo \cite{asmo_zenodo}.

\subsubsection{Ontology design, evaluation and reuse of existing ontologies}

The ontology engineering process followed a hybrid, iterative methodology combining bottom-up knowledge acquisition from heterogeneous data sources with domain expert-driven conceptual modeling. The ontology was designed in a modular fashion, with separate but connected components (e.g., CMSO, ASMO, and defect modules), enabling extensibility. The development approach is closely aligned with the principles of the NeOn methodology \cite{Suarez-Figueroa2012}, particularly in its emphasis on reuse, modularity, and iterative refinement.

Terminology was derived from non-ontological sources, primarily atomistic simulation software and materials databases, while relationships were formalized based on domain expertise. The resources used for terminology extraction are: pyscal \cite{Menon2019}, pyironatomistics \cite{Janssen2019}, ASE \cite{HjorthLarsen2017}, pymatgen \cite{Ong2013}, atomman\cite{Hale2018}, LAMMPS \cite{THOMPSON2022}, VASP \cite{Kresse1996}, QuantumEspresso \cite{Giannozzi2009}, Materials Project \cite{Jain2013}, OPTIMADE \cite{Evans2024}, The Crystallographic Information File (CIF) \cite{Hall1991}.

Existing ontologies, including PROV-O \cite{provo}, QUDT \cite{qudt_fairsharing}, and MDO \cite{Li2020}, were reused where appropriate to promote interoperability.

\begin{itemize}
    \item PROV-O: reused mainly in ASMO to describe provenance around simulations and workflows. ASMO aligns simulation processes with prov:Activity and uses PROV concepts such as entities, agents, and relations like prov:used, prov:wasGeneratedBy,\\ and prov:wasAssociatedWith to capture how calculations are performed, by whom, and with which software or inputs.

    \item QUDT: reused in ASMO and CMSO for unit handling. It provides the unit layer for physical quantities and calculated properties, allowing values such as energy, pressure, length, temperature, and time to be associated with explicit units through classes such as qudt:Unit.

    \item MDO: reused in ASMO for materials modelling concepts, especially electronic-structure method details such as exchange-correlation functional families. This gives ASMO a way to reference standard method concepts without redefining them.
\end{itemize}

The ontology was evaluated using a combination of criterion-based, task-based, and structural validation approaches. First, requirement analysis and competency questions were defined to guide the development and assess whether the ontology can support relevant scientific queries; a subset of these competency questions is provided in the Supplementary Information. Second, the ontology was validated in application-driven scenarios by integrating and querying heterogeneous datasets. This evaluation ensures that the ontology captures the structure and semantics of real-world materials science data. Finally, structural consistency was assessed using automated tools such as the Ontology Pitfall Scanner \cite{Poveda2014}, and logical consistency can be verified using OWL reasoners such as HermiT \cite{Glimm2014}, ensuring the absence of common modeling pitfalls and logical inconsistencies.

\subsection{Software infrastructure for ontology-based simulation data annotation}\label{sec:atomrdf}

Although ontology-based representations offer clear advantages for interoperability and machine interpretability, their direct use through the Resource Description Framework (RDF) \cite{rdf11-concepts} and the Web Ontology Language (OWL) \cite{owl2_2012} remains challenging in routine scientific workflows \cite{Tudorache2020}. In practice, atomistic simulation data and metadata are generated within heterogeneous software environments and file formats, where direct interaction with RDF or OWL is neither natural nor efficient. Our software architecture is therefore designed to overcome this barrier by introducing an intermediate, ontology-aligned representation of metadata. This representation retains the semantic structure required for consistent graph construction, while exposing them through data structures that are familiar, lightweight, and directly usable within existing scientific software environments.

\begin{figure}[ht!]
\centering
\includegraphics[scale=0.21]{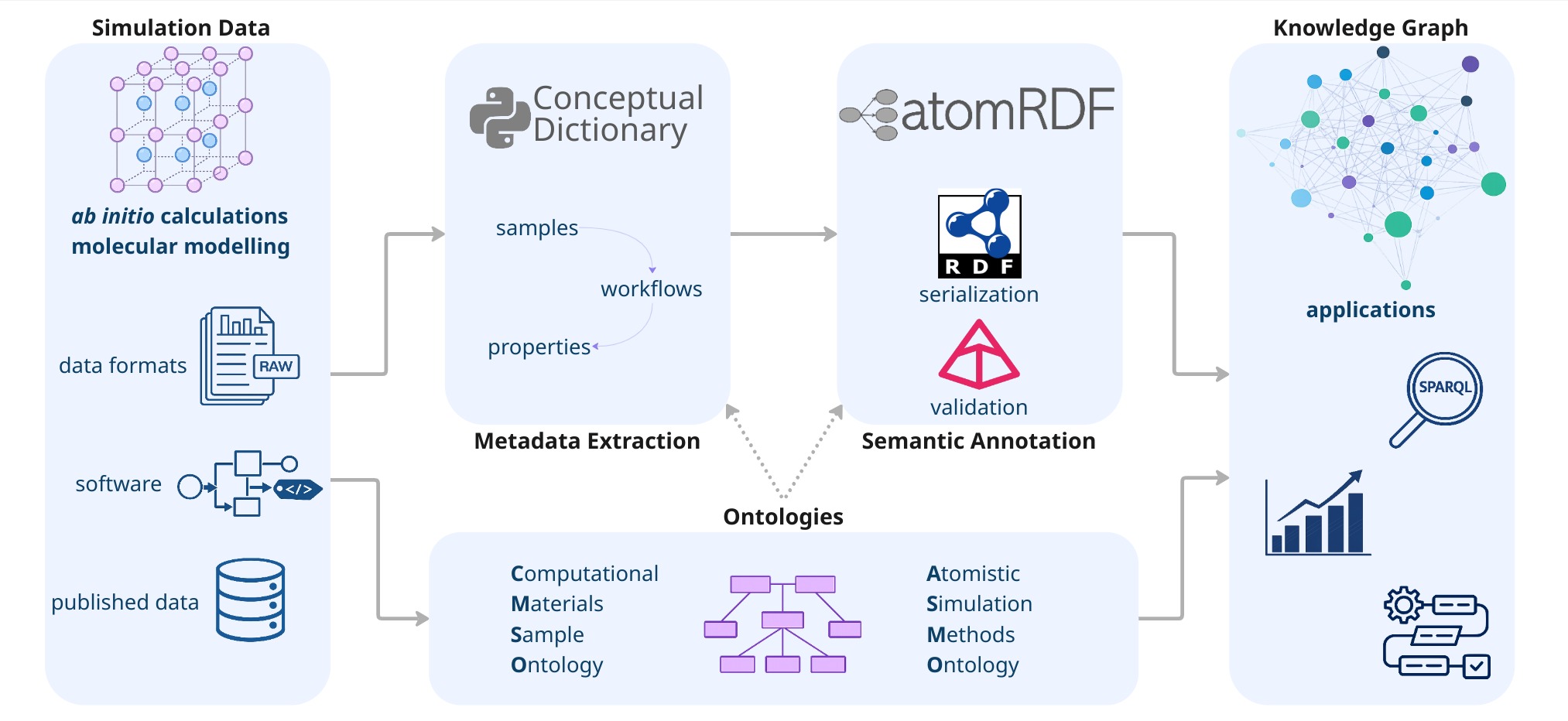}
\caption{Overview.} 
\label{fig:overview}
\end{figure}

On this basis, the infrastructure is organized not as a monolithic graph creation system, but as a layered pipeline from structured metadata capture to knowledge graph representation. By separating metadata acquisition, semantic modeling, and graph construction, the architecture enables different user groups and software components to interact with the system at the level most appropriate to them, while maintaining semantic consistency across the full stack. In practice, this means that the same framework can support manual annotation of heterogeneous data as well as automated integration into simulation workflows. An overview of the framework can be seen in Figure \ref{fig:overview}.

\subsubsection{Conceptual metadata capture}

The first layer is conceptual metadata capture, implemented through \texttt{conceptual\_dictionary}. It provides reusable, ontology-aligned metadata templates derived from the concepts introduced in Sec.~\ref{sec:onto}, exposing them in a form suitable for practical use in scientific workflows. These templates are available in common human- and machine-readable formats such as \texttt{YAML} and \texttt{JSON}, as well as through an importable Python dictionary interface with built-in validation for controlled fields.
The conceptual dictionary thus provides an entry point for metadata in many ways: the templates can be filled manually, but more importantly they can be incorporated into software tools, automated workflows, and workflow managers without significant overhead in terms of software imports or dependencies. 
From a user perspective, no interaction with RDF representations is needed. In this way, metadata are captured at source in a structured, ontology-consistent form while remaining compatible with existing simulation environments.
This provides a practical route for handling existing and heterogeneous data sources, since dedicated parsers can be designed to populate the templates from legacy file formats, databases, or extracted records. In addition, the lightweight and explicit structure of the templates makes them a suitable interface for LLM-based knowledge extraction pipelines, where extracted entities and relations can first be normalized in a controlled metadata representation before being transformed into RDF \cite{Xu2024}.

The templates follow a small set of core abstractions reflecting the structure of atomistic simulations: \texttt{computational\_sample}, \texttt{workflow}, and \texttt{math\_operations}. The \texttt{computational\_sample} section captures the definition of each material system, including a unique identifier. The \texttt{workflow} section represents simulation activities, such as molecular dynamics or density functional theory calculations. \texttt{math\_operations} encode simple post-processing steps used to derive material properties, enabling derived quantities to be captured explicitly together with their computational provenance. 
Together, these abstractions provide a lightweight, semantically guided representation of simulation inputs, outputs, and intermediate processing steps prior to graph construction.

\subsubsection{Ontology-aligned data models in atomRDF}

\texttt{atomRDF} implements data models derived from the conceptual layer and aligned with the ontology suite. 
Within the software architecture, \texttt{atomRDF} acts as the central translation layer between lightweight metadata representations and ontology-grounded graph objects.
Pydantic \cite{Pydantic2026} data classes are used to provide typed, ontology-aligned representations of the main entities and relations captured in the metadata layer. 
\texttt{atomRDF} includes parsers that can read \texttt{conceptual\_dictionary} serializations in \texttt{YAML} or \texttt{JSON}, which are then used to populate the corresponding data classes, providing the transition from metadata capture to ontology-grounded software objects. 
These data models offer strong validation, ensuring the quality and internal consistency of data before graph construction. 
They act as the transformation boundary between structured Python/\texttt{JSON} objects and RDF graph representations, enabling ontology-consistent annotation in scientific workflows.

Each data model in \texttt{atomRDF} has a strong connection to the ontologies, including persistent identifiers and ontology-grounded mappings for each attribute.
They also have built-in \texttt{from\_graph} and \texttt{to\_graph} methods. The \texttt{to\_graph} methods serialize the information held in the data class into RDF triples, while the \texttt{from\_graph} methods reconstruct Python objects from an existing graph representation. This bidirectional conversion is a key architectural feature, as it enables the same data model to support both graph construction and reuse of graph data within scientific workflows, in downstream applications.

\texttt{atomRDF} further includes a \texttt{KnowledgeGraph} object inheriting from an \texttt{rdflib} \cite{RDFLib2025} graph. Data serialized from the ontology-aligned models can be added directly into the knowledge graph. 
In this way, the knowledge graph is not built through \textit{ad hoc} triple generation, but through ontology-aligned software objects that preserve the structure, semantics, and provenance of the original metadata. This also means that users can interact with the system at different abstraction levels: through template files, validated Python data models, or directly through the graph representation, depending on the requirements of the application.

A consequence of this design is that ontology engineering and software interfaces remain coupled without being directly combined: the ontology defines the formal meaning, while the software layers provide representations that are practical for scientists and workflows to adopt. This design further shifts validation from \textit{post hoc} graph checking to preemptive constraint enforcement during data generation. Because knowledge graph instances are created through controlled, ontology-aligned data models, malformed or incomplete data can be detected before serialization, improving consistency at source. While this differs from SHACL-based validation of arbitrary RDF graphs, it provides comparable guarantees within the scope of the present infrastructure. Future extensions may expose these constraints as SHACL shapes to support validation of externally integrated RDF data.

\subsection{FAIR alignment and sustainability}

The ontologies are developed within the context of NFDI-MatWerk, the German national data infrastructure consortium for materials science and engineering, providing an institutional framework for long-term governance and community-driven development. The infrastructure is used across multiple research groups in domain-specific use cases. Development is conducted openly via GitHub, enabling community contributions through issues and pull requests, while versioned releases are archived on Zenodo. Sustainability is supported through a modular design that allows the ontology to evolve alongside emerging scientific requirements, ensuring both maintenance and adoption within an active research community.

The presented infrastructure is designed to support the FAIR (Findable, Accessible, Interoperable, Reusable) principles across multiple layers, including the ontology, knowledge graph, data, workflows, and software stack. By integrating semantic modeling with controlled data generation and provenance capture, the system enables FAIR-by-design representation of materials simulation data.

\begin{itemize}
    \item Findability: All entities in the knowledge graph are identified by globally unique and persistent IRIs defined by the ontology. Instance-level data are assigned UUID-based identifiers, while structure representations are hash-based, ensuring that identical structures are uniquely and consistently referenced across datasets. The knowledge graph is published via Zenodo with versioned releases and is accessible through a SPARQL endpoint \cite{zenodo_data, github_data}, enabling structured and queryable access. The use of a well-defined ontology further supports semantic discoverability of entities such as materials, properties, and workflows.
    \item Accessibility: The ontology and knowledge graph are openly available via Zenodo, with version control and development hosted on GitHub. Access is provided through both direct download and SPARQL querying, ensuring flexible and programmatic data retrieval. In addition, deployment via a terminology service enables API-based access to ontology terms. All resources are openly accessible and versioning ensures long-term availability and reproducibility.
    \item Interoperability: achieved through the use of standard semantic web technologies (RDF, RDFS, OWL) and the reuse of established ontologies. Entities are further linked to external resources such as ChEBI \cite{chebi} for chemical elements and Wikidata \cite{wikidata} for crystal structure information, enabling cross-domain integration. Workflows are explicitly represented as semantic graphs, allowing seamless integration of data, methods, and provenance across heterogeneous sources.
    \item Reusability: ensured through rich metadata and explicit provenance modeling. The knowledge graph captures detailed information on simulation conditions, methods, software, and input/output relationships, enabling users to interpret and reproduce results. Provenance links to original datasets, including references to DOIs and authors, are preserved, ensuring traceability. Furthermore, all components, including ontology, data, and software, are released under open licenses, promoting reuse and adaptation. By representing both data and workflows in a unified framework, the infrastructure enables data reuse across different scientific contexts.
\end{itemize}

\section{Demonstration of interoperability and data reuse}

To demonstrate interoperability and data reuse, we integrated atomistic simulation data from heterogeneous sources, including Zenodo archives, supplementary information associated with publications, and Git-based repositories from these publications: \cite{Tschopp2015, Mai2025, Chen2025, chen_jiahao2025, Choi2025, Brink2023, ZHENG2020}. 
These records differed not only in format and level of structure, but also in the completeness of their metadata and provenance. Where required, the datasets were complemented with contextual information extracted from the corresponding publications in order to recover missing simulation details and computational history. The collected records were then normalized through a combination of manual annotation and automated parsers, first into \texttt{conceptual\_dictionary} templates and subsequently into the knowledge graph using \texttt{atomRDF}. In this way, the integration step itself serves as a direct demonstration of the infrastructure: heterogeneous records are transformed into a common, ontology-aligned representation that supports complex querying and data reuse.

To further demonstrate the flexibility of the infrastructure, we supplemented the aggregated datasets with newly generated data. Density functional theory (DFT) calculations were performed using the PBE exchange–correlation functional \cite{Perdew1996} as implemented in VASP \cite{Kresse1996} to compute vacancy formation energies for eight elemental systems. In addition, formation energies of substitutional and interstitial He defects were calculated for the same systems. All calculation parameters and workflows are captured and available in the knowledge graph. These calculations were parsed after completion using our parser workflows to generate \texttt{conceptual\_dictionary} templates and then added to the knowledge graph. 

To also illustrate a fully automated data generation and annotation pipeline, we carried out vacancy formation energy calculations for the same elemental systems using GRACE-2L-OMAT \cite{Lysogorskiy2026}, a universal machine-learning potential. In this case, automated Python workflows not only executed the calculations but also generated the corresponding metadata templates directly, without \textit{post hoc} manual intervention.

\begin{figure}[ht!]
\centering
\includegraphics[scale=0.15]{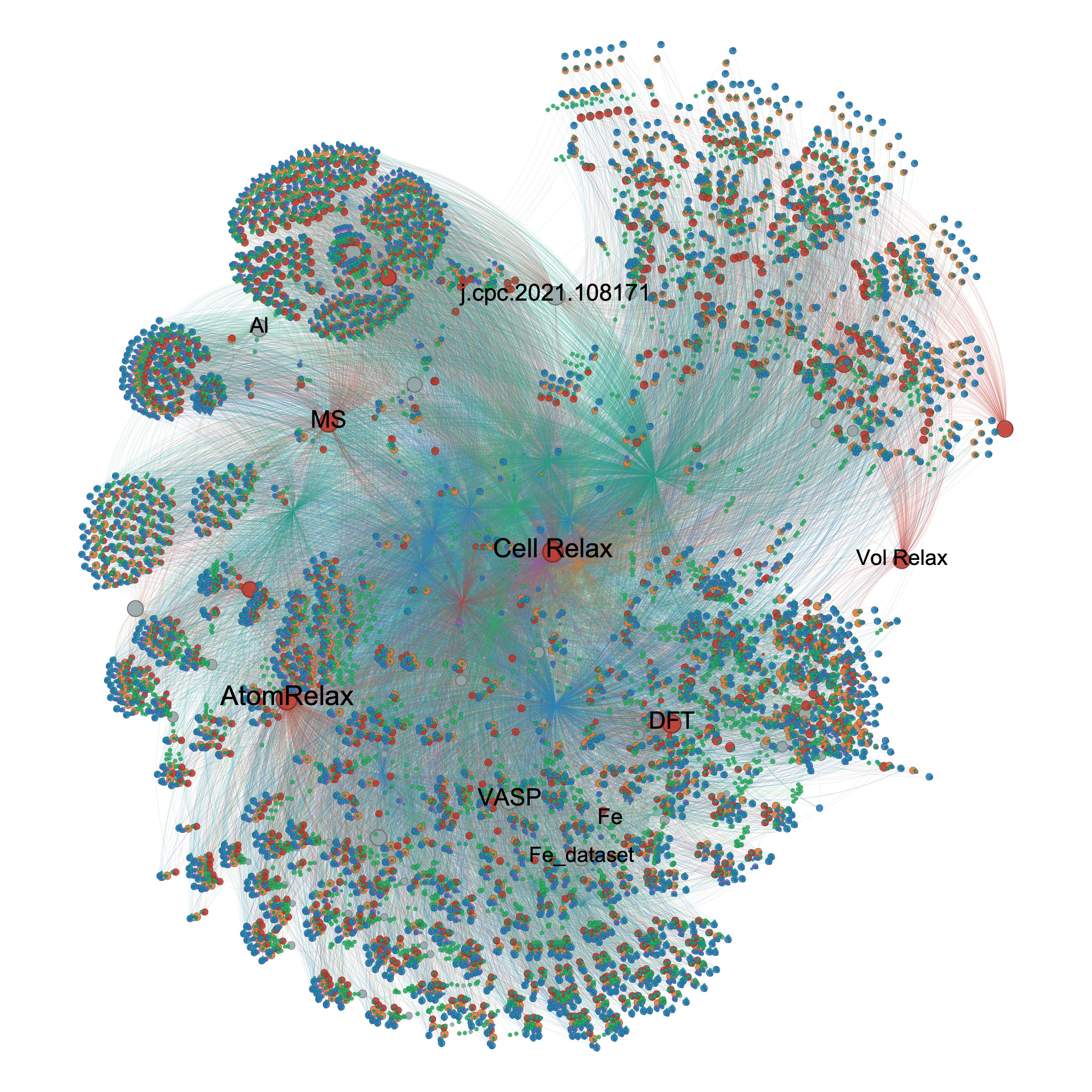}
\caption{Visualisation of the knowledge graph of atomistic simulation data. Nodes represent RDF resources such as simulation samples, crystal structures, chemical elements, calculated properties, interatomic potentials, and computational methods; coloured by semantic category (orange: Sample; blue: Structure; teal: Property; red: Calculation; green: Element; purple: Material; gold: Potential). Labels are shown only for the ten most-connected nodes. The visualized data is subsampled from the knowledge graph (13\% of the full graph).}
\label{fig:kg}
\end{figure}

The resulting knowledge graph used in the present work contains 757{,}253 triples describing 7{,}926 computational samples. The structure of the resulting knowledge graph, including the main entity types and their connectivity, is illustrated in Fig.~\ref{fig:kg}. Although not all currently supported data models are represented in this graph, the framework already includes models for a wider range of methods and simulation outputs, including equation-of-state calculations, free-energy calculations using the quasiharmonic approximation or thermodynamic integration \cite{Menon2021}, and additional defect classes such as stacking faults and dislocations. The present graph therefore represents both a concrete integrated resource and a subset of a broader extensible infrastructure for atomistic simulation data.

\subsection{Semantic integration of heterogeneous atomistic simulation data}

We first demonstrate how the knowledge graph enables efficient exploration of heterogeneous atomistic simulation data in a unified semantic space. We focus on grain boundary data, which provide a demanding test case because the relevant descriptors span composition, grain boundary character, simulation methodology, and target property \cite{sutton1996interfaces}. Such data are particularly heterogeneous because grain boundaries are defined in a five-degree-of-freedom configurational space and are typically distributed across independent datasets with inconsistent structure and terminology. This makes grain boundary data a stringent test of whether semantically aligned integration can support retrieval and comparison across sources.

\begin{figure}[h!]
\centering
\includegraphics[scale=0.7]{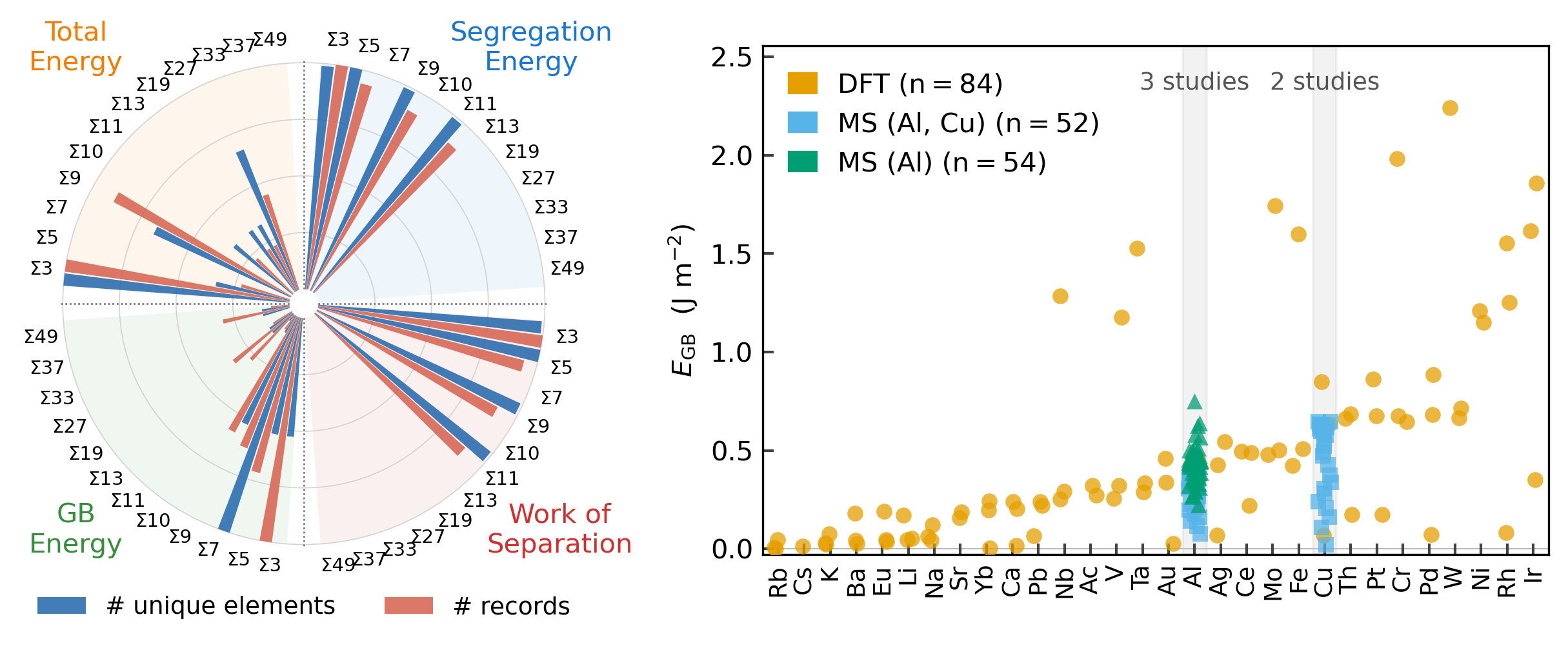}
\caption{Semantic integration of heterogeneous grain boundary data in the knowledge graph. The left panel summarizes the currently available grain-boundary-related properties across elements, methods, and data sources. The right panel shows the result of a targeted query for the energy of $\Sigma$3 grain boundaries calculated with different methods (n is the total number of data records), illustrating how semantically aligned records can be retrieved across datasets to reveal existing coverage and gaps in the current data landscape.}
\label{fig:gb}
\end{figure}

Fig.~\ref{fig:gb}a summarizes the grain boundary data currently represented in the knowledge graph. The integrated data include four major classes of calculated properties: total energy, grain boundary energy, segregation energy, and work of separation, each available for different $\Sigma$ values. In the coincidence site lattice formalism \cite{Randle1997}, $\Sigma$ denotes the reciprocal density of coincident lattice sites between the two adjoining crystals and is therefore widely used to classify grain boundary types. These properties are available for multiple elements and grain boundary types, and are associated with different simulation methods and interatomic potentials. The figure therefore highlights not only the volume of available records, but also the breadth of heterogeneous data that have been normalized into a common representation. In this sense, the knowledge graph functions not merely as a repository, but as an integration layer in which otherwise disconnected data become jointly searchable and comparable.

An advantage of this representation is the ability to perform targeted queries over semantically-aligned classes rather than over source-specific file structures or naming conventions. For example, a query restricted to $\Sigma$3 grain boundaries can retrieve all matching entries across datasets, independent of their original source, file structure, or naming convention. The results of such a SPARQL query are shown in the right panel of Fig.~\ref{fig:gb}. This query shows that grain boundary energies have been calculated for $\Sigma$3 grain boundaries for about 30 elements using DFT. In contrast, more detailed studies, for example using molecular dynamics, are available only for a subset of these systems, in particular for Al and Cu. Such queries make it possible to assemble a structured overview of available data for a scientifically meaningful question, including which elements, properties, methods, and potentials are already covered. These targeted queries therefore provide a machine-actionable overview of existing studies for a given grain boundary and can serve as a starting point for planning new calculations or identifying under-explored systems and data gaps. Because the atomic configurations themselves are also represented in the graph and linked to their provenance, the underlying structures remain reproducible and reusable.

\begin{figure}[ht!]
\centering
\includegraphics[scale=0.8]{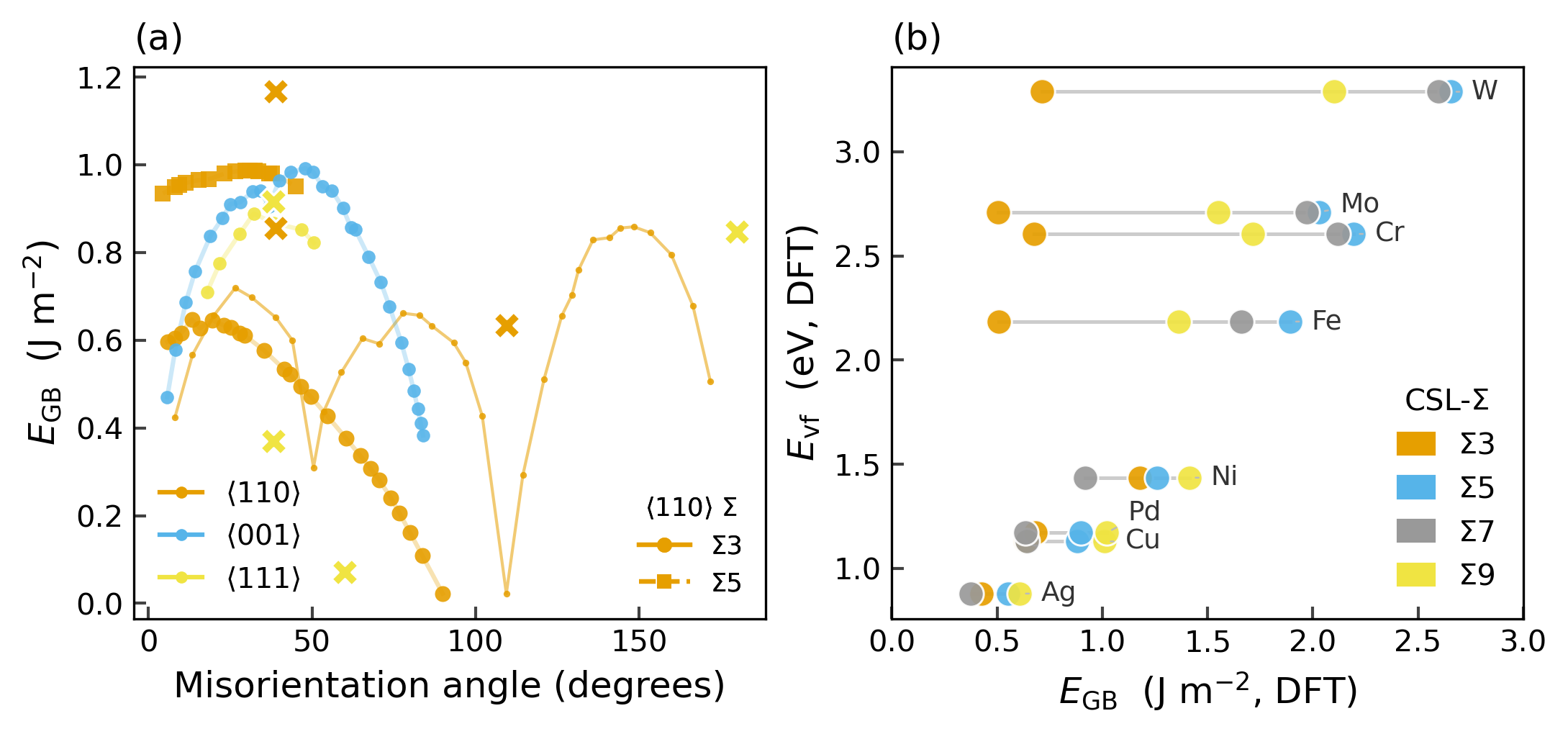}
\caption{Cross-dataset grain boundary properties queried from the knowledge graph.
(a) Grain boundary energy ($E_\mathrm{GB}$) versus misorientation angle for Cu along the
$\langle 110 \rangle$, $\langle 001 \rangle$, and $\langle 111 \rangle$ tilt axes.
Continuous curves are from molecular dynamics, while DFT reference values are marked
with crosses. Within $\langle 110 \rangle$, $\Sigma3$ (solid) and $\Sigma5$ (dashed)
boundaries are highlighted, with further data shown as dots.
(b) Correlation between DFT vacancy formation energy ($E_\mathrm{vf}$) and DFT median
$E_\mathrm{GB}$ for eight transition-metal elements across four $\Sigma$ values.}
\label{fig:gb2}
\end{figure}

We further demonstrate how the knowledge grap infrastructure can be used not only to retrieve heterogeneous data, but also to identify scientific trends across datasets. In Fig.~\ref{fig:gb2}(a), we plot grain boundary energy as a function of misorientation angle, resolved by $\Sigma$. The resulting distributions recover the well-known energetic minima associated with special low-$\Sigma$ boundaries, indicating that the integrated data reproduce established crystallographic trends and are therefore internally consistent \cite{WOLF1989}. This illustrates that semantically integrated data can support not only retrieval, but also physically meaningful comparative analysis across previously separate studies.

A second example is shown in Fig.~\ref{fig:gb2}(b), where we relate vacancy formation energy to grain boundary energy by combining quantities from different datasets. This illustrates an important feature of the infrastructure: once represented in a common graph, independently generated data can be queried together to test cross-property relationships. In our case, the data suggest that, across elements and for higher-$\Sigma$ boundaries, vacancy formation energy and grain boundary energy are positively correlated. While this trend should be interpreted cautiously in view of the available data coverage, it shows how the knowledge graph can support the discovery of physically meaningful relations that would be difficult to identify from isolated datasets alone.

\subsection{Extracting thermodynamic properties from existing data} \label{sec_thermal_expansion}

We further demonstrate the value of representing simulation data as a knowledge graph by deriving physical quantities that were not explicitly reported in the original datasets. This is particularly relevant because large volumes of simulation data are routinely generated in individual studies, while many potentially useful derived quantities remain uncomputed and therefore remain effectively hidden in existing data collections.

As an example, we query the knowledge graph for the volume per atom of pure elemental systems obtained from molecular dynamics (MD) simulations performed in the isothermal--isobaric ($NPT$) ensemble. To ensure comparability across the retrieved data, we restrict the query to cubic systems and to simulations carried out using the same interatomic potential. Such a selection would be difficult to perform reliably from structure files alone, and remains cumbersome even in conventional workflow-based environments, where the relevant metadata are often distributed across files, scripts, or publications. In contrast, the graph representation enables direct querying across structures, simulation conditions, and provenance metadata in an integrated manner.

\begin{figure}[ht!]
\centering
\includegraphics[width=0.5\textwidth]{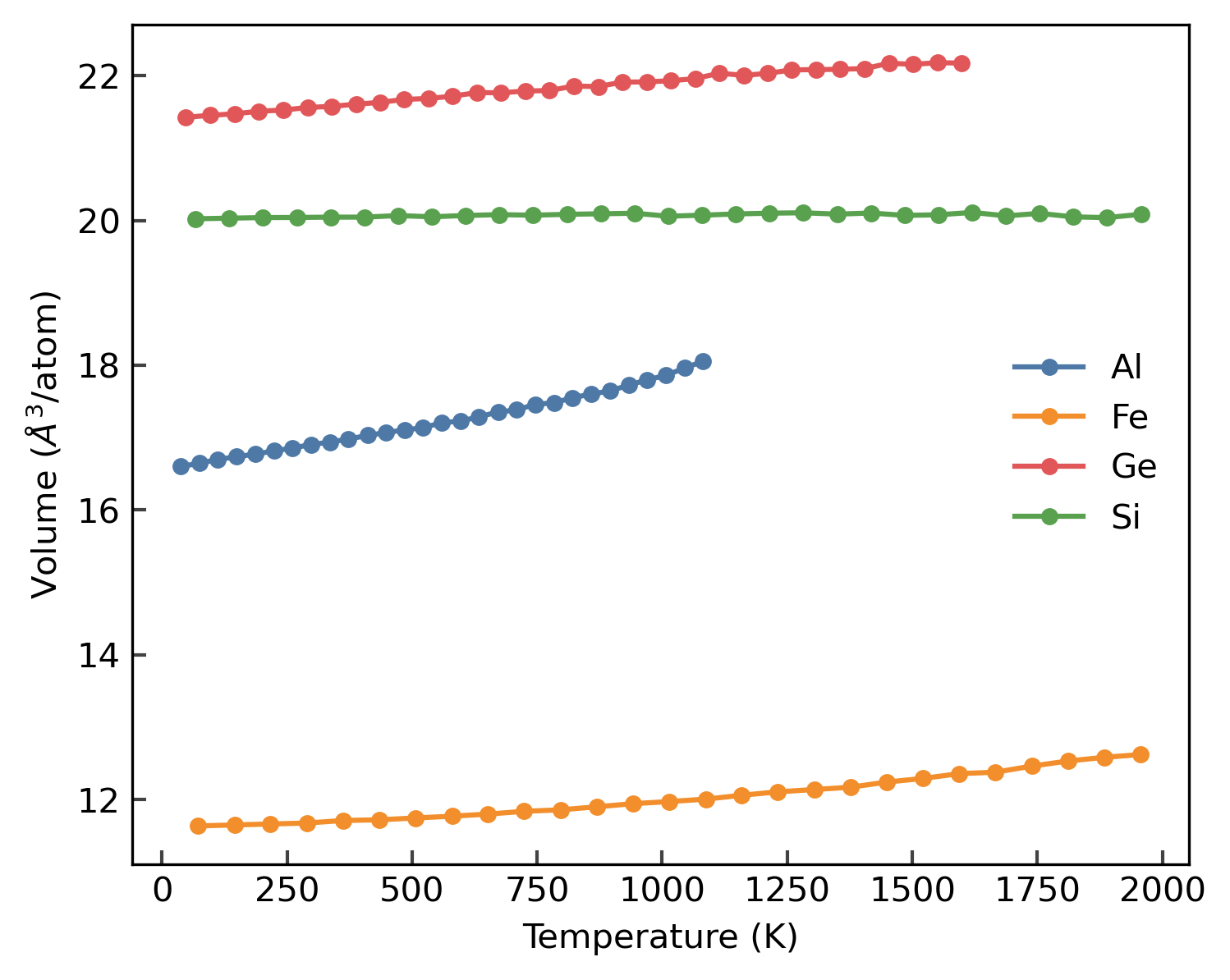}
\caption{Volume per atom as a function of temperature for selected elemental systems retrieved from the knowledge graph. By querying structurally and methodologically comparable molecular dynamics data, the volumetric thermal expansion coefficient can be estimated from the slope of the resulting volume-temperature relation.}
\label{fig_thermal_expansion}
\end{figure}

The queried results are shown in Fig.~\ref{fig_thermal_expansion}, and the corresponding queries and analysis notebooks are provided as part of the accompanying resources. From these data, we extract the volumetric thermal expansion coefficient, a fundamental thermodynamic property that quantifies the change in volume with temperature, defined as: $\alpha = \frac{1}{V} \frac{\partial V}{\partial T}$. Applying this equation to the queried data yields volumetric thermal expansion coefficients ranging from $4\times 10^{-6}$ K$^{-1}$ for Si to $90\times 10^{-6}$ K$^{-1}$ for Li, with intermediate values of $68\times 10^{-6}$ K$^{-1}$ for Al, $22\times 10^{-6}$ K$^{-1}$ for Fe, and $25\times 10^{-6}$ K$^{-1}$ for Ge.
Inspection of the retrieved records shows that the data used in this example originate primarily from the DC3 dataset~\cite{Chung2022}. That dataset was originally created to support the development of machine-learning models for identifying crystal structures in MD simulations and contains atomic configurations extracted from selected simulation timesteps at different temperatures for several systems, including eight elemental materials. To reuse such data for a property calculation of this kind, one would typically need to combine the atomic configurations with additional information from the associated publication, such as the simulation conditions and the interatomic potential employed. By integrating these metadata into the knowledge graph, these connections become directly queryable, enabling the extraction of derived physical quantities from pre-existing simulation data. This use case illustrates how the proposed framework supports the reuse of existing datasets to generate new scientific value.


\subsection{Ontology-enabled inference and provenance over simulation workflows}

A central requirement for reusable simulation data is that the computational history, or provenance, of any stored result remains accessible long after the original calculation was performed. In conventional practice, this information is distributed across input files, submission scripts, output files, and methods sections of publications, requiring substantial manual cross-referencing to determine which structures, simulation codes, parameters, and post-processing steps were combined to calculate a given material property. In computational materials science, this is particularly important because calculated property values can depend strongly on the methodology and sequence of steps used.

Our infrastructure allows workflows to be annotated directly at source and therefore enables capture of complete provenance in the knowledge graph. This supports not only reproducibility, but also interpretability and reuse. In this sense, our approach enables a two-way provenance model: it can preserve provenance forward from the point of calculation, but can also recover provenance backward from existing results and reconnect them to the computational context from which they originated. We demonstrate this using vacancy formation energy, a prototypical property calculation in materials simulation. Even such a standard calculation can involve multiple structures, simulation steps, files, scripts, and post-processing operations \cite{Walsh2024}. We perform vacancy formation energy calculations for eight elements using both density functional theory (DFT) and molecular statics (MS) with the foundational GRACE model \cite{Lysogorskiy2026}, and annotate them at source using the conceptual dictionary. These annotated calculations are then included in the knowledge graph.

\begin{figure}[ht!]
\centering
\includegraphics[scale=0.19]{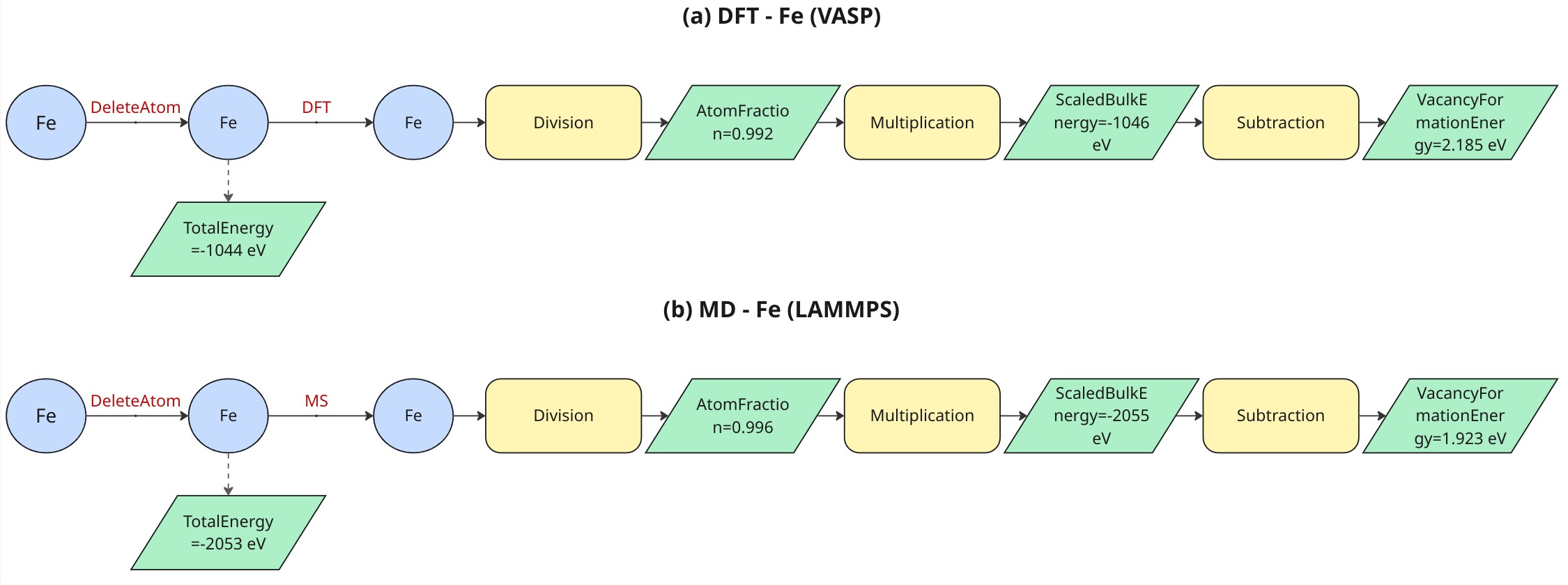}
\caption{Machine-readable provenance representation for vacancy formation energy calculations, comparing workflows based on DFT and molecular statics. Although the underlying simulation methods differ, the graph representation reveals their shared workflow structure and makes both common and method-specific steps explicit.}
\label{fig:prov}
\end{figure}

We then explore the provenance of the calculated property using atomRDF:

\begin{verbatim}
prov = kg.trace('property:formationenergy')
prov.visualize()
\end{verbatim}

This creates a machine-readable provenance record together with a visualization, as shown in Fig.~\ref{fig:prov}, which compares the generated provenance for the DFT and molecular statics calculations. An important result is that the workflows are immediately recognizable as structurally equivalent, even though the underlying simulation methods differ. This makes explicit that one method can, in principle, be substituted for another within the same overall workflow pattern, and also enables direct comparison of the calculated values. More generally, the graph representation makes methodological commonality explicit across computational approaches that are usually documented separately. As a result, provenance is not only preserved, but normalized into a form that can be queried, compared, and interpreted across methods.

Most importantly, the provenance captures the post-processing operations required to obtain the final vacancy formation energy. In particular, it records that the energy must be scaled with respect to the number of atoms and that an energy difference between a defective sample and a reference sample must be evaluated to obtain the target property. These operations are often carried out manually in post-processing scripts or notebooks and are rarely recorded in a structured form, leading to a break in provenance. Our approach closes this gap by representing these derived operations explicitly in the graph.

We further demonstrate the machine-readability of the provenance, and highlight the remaining missing links, by attempting to reconstruct the workflow automatically from the knowledge graph. This is done through a \texttt{reconstruct} method, which generates a directory containing recreated structures in ASE JSON format together with a Python file describing the workflow steps. The original workflow engine or execution scripts are not stored in the knowledge graph. Instead, executable code is recreated by filling code templates with inferred input parameters and workflow steps extracted from the graph.

We test this approach by reconstructing the molecular statics workflow for the vacancy formation energy calculation. Although the knowledge graph captures the identity of the interatomic potential employed, the exact way in which that potential is specified in LAMMPS, for example through \texttt{pair\_style} and \texttt{pair\_coeff} commands, as well as the specific potential file itself, are not yet available. This reflects a broader issue in the field, since interatomic potential files are generally not version controlled or persistently linked to the simulation record. Although there are important efforts in this direction, such as OpenKIM and repositories for interatomic potentials \cite{Tadmor2011, Hale2016}, a general solution is still lacking. At the same time, this example shows that the graph makes such missing information explicit, enabling identification of the workflow stages that remain outside current FAIR and reproducibility practices.

Once the potential-specific information is provided, the automatically generated workflow becomes fully executable and is able to reproduce the exact results. Such two-way provenance is, in our view, a key step towards making simulation data not only findable and reusable, but also computationally reconstructible.

\section{Discussion and conclusions}

In this work, we have developed an ontology-based infrastructure for atomistic simulation data, consisting of domain ontologies and a software stack for knowledge graph construction. The approach enables metadata to be captured both from existing published datasets and directly from simulation workflows at the point of data generation. These data are normalized through a lightweight metadata layer and transformed into ontology-aligned knowledge graph representations, allowing structured, machine-readable access to simulation data and provenance.

\paragraph{Impact} The main impact of this work is the demonstration of interoperable integration of heterogeneous atomistic simulation data across sources, methods, and formats. Through the ontology layer, semantic inconsistencies between datasets are resolved and data are brought into a common representation, enabling consistent interpretation and reuse. The infrastructure further provides a practical bridge between ontology-based representations and scientific workflows, allowing metadata to be captured at source while remaining usable within existing computational environments.

We demonstrate three core capabilities:
\begin{itemize}
    \item This enables cross-dataset querying and comparative analysis in complex material systems, as demonstrated for grain boundary data. 
    \item It further allows reuse of existing data to generate new scientific insights, including the extraction of physical quantities that were not explicitly reported in the original studies.
    \item It enables two-way provenance: workflows can be tracked forward from data generation and also reconstructed backward from existing results. In this way, the framework moves beyond data reuse towards computational reproducibility through partial workflow reconstruction.
\end{itemize}

More broadly, the approach supports the FAIR principles by improving findability, interoperability, and reuse of simulation data, while also advancing reproducibility through explicit representation of workflows and derived operations. At the same time, the use of ontology-aligned metadata templates provides a practical route towards standardization of simulation metadata across heterogeneous sources.

\paragraph{Limitations} The approach depends on the availability and quality of input metadata, which can be incomplete or inconsistent for legacy datasets. External dependencies, such as interatomic potential files and software-specific input configurations, are not yet fully captured in a structured form. Workflow reconstruction is currently limited by the absence of standardized representations of simulation engines and execution details, and integration of heterogeneous data still requires manual curation or the development of dedicated parsers. The demonstrated knowledge graph does not yet cover the full range of simulation methods and properties relevant to materials science, and scalability for very large or continuously evolving datasets remains to be explored. Validation is currently focused on controlled data generation within the framework and does not yet extend to arbitrary external RDF data. More generally, integration into broader, aligned ontology ecosystems for provenance and upper-level semantics remains an open challenge, as highlighted by recent work on aligning provenance ontologies such as PROV-O with upper ontologies \cite{Prudhomme2025}. Long-term impact will depend on community-driven governance, extension, and alignment of the underlying ontologies.

\paragraph{Future plans} Future work will extend the ontology and data models to cover a broader range of simulation methods, properties, and defect types, and to better represent the multiscale nature of materials across different length scales. Workflow reconstruction capabilities will be developed further towards fully executable and reproducible simulation pipelines, including improved representation of external dependencies and execution details. Automated metadata extraction from literature and legacy datasets will also be explored to support scalable population of the knowledge graph, including the use of LLM-based approaches in combination with structured representations \cite{Buehler2024}. More broadly, ontology-aligned simulation data provide a basis for integration with data-driven and machine learning methods. Continued development will focus on strengthening interoperability through alignment with external ontologies and standards, as well as supporting community adoption and contribution to the ontology and software ecosystem.

\section*{Data Availability}

The complete set of RDF triples generated from the datasets used in this work, comprising the knowledge graph is provided via Zenodo \cite{zenodo_data}.  
In addition, a demonstrator for exploring the knowledge graph is available online \cite{github_data}.

\section*{Code Availability}

All software developed in the context of this work, including \texttt{conceptual\_dictionary} and \texttt{atomRDF}, is publicly available through their respective Git repositories \cite{Menon_conceptual_dictionary_2026, Menon_atomRDF_2026}.  
The workflows used to analyse the knowledge graph and generate the trends and other results reported in this work are also available in a public repository \cite{github_data}.

\section*{Acknowledgements}
This work is supported by the consortium NFDI-MatWerk, funded by the Deutsche Forschungsgemeinschaft (DFG, German Research Foundation) under the National Research Data Infrastructure – NFDI 38/1 – project number 460247524.

\bibliographystyle{simplejournal}
\bibliography{references}

\end{document}